\documentclass[aps,groupedaddress,10pt,twocolumn,amsmath,amssymb]{revtex4}
\usepackage[dvips]{graphics}
\usepackage{epsfig}
\usepackage{dcolumn}
\usepackage{xspace}
\usepackage{amssymb,amsmath}
\begin{document}

\title{On how good DFT exchange-correlation functionals are for H bonds in small water clusters: Benchmarks
approaching the complete basis set limit}

\author{Biswajit Santra$^1$ }
\author{Angelos Michaelides$^{1,2}$}
\email{michaeli@fhi-berlin.mpg.de}
\author{Matthias Scheffler$^1$}
\affiliation{$^1$Fritz-Haber-Institut der Max-Planck-Gesellschaft, Faradayweg 4-6, 14195 Berlin, Germany \\
$^2$Virtual Materials Laboratory, London Center for Nanotechnology
and Department of Chemistry, University College London, London WC1E
6BT, UK}

\begin{abstract}
The ability of several density-functional theory (DFT)
exchange-correlation functionals to describe hydrogen bonds in small
water clusters (dimer to pentamer) in their global minimum energy
structures is evaluated with reference to second order M\o ller
Plesset perturbation theory (MP2). Errors from basis set
incompleteness have been minimized in both the MP2 reference data
and the DFT calculations, thus enabling a consistent systematic
evaluation of the true performance of the tested functionals. Among
all the functionals considered, the hybrid X3LYP and PBE0
functionals offer the best performance and among the non-hybrid GGA
functionals mPWLYP and PBE1W perform the best. The popular BLYP and
B3LYP functionals consistently underbind and PBE and PW91 display
rather variable performance with cluster size.
\end{abstract}

\maketitle

%
%
\textbf{I. Introduction}

Density-functional theory (DFT) is the most popular theoretical
approach for determining the electronic structures of polyatomic
systems. It has been extensively and successfully used to tackle all
sorts of problems in materials science, condensed matter physics,
molecular biology, and countless other areas. Many of these studies
have involved the treatment of systems containing hydrogen bonds.
Hydrogen bonds are weak (10-30 kJ/mol $\approx$ 100-300 meV/H bond)
bonds of immense widespread importance, being the intermolecular
force responsible for holding water molecules together in the
condensed phase, the two strands of DNA in the double helix, and the
three dimensional structure of proteins \cite{Jeffrey_book}. A
particularly important class of H-bonded systems are small water
clusters.
Small water clusters have been implicated in a wide range of
phenomena (for example, environmental chemistry and ice nucleation
\cite{saykally_science_review,michaelides_morgenstern}) and,
moreover, are thought to provide a clue as to the properties of
liquid water. However the ability of DFT to quantitatively describe
H bonds between H$_2$O molecules in either small water clusters or
the liquid state remains unclear. This is particularly true in light
of recent experimental and theoretical studies which have raised
concerns over the ability of DFT to reliably describe the structure
and properties of liquid water
\cite{Todorova,artacho,tuckerman,parrinello_water,
nilsson_science,saykally_science_2004,grossman_water}. \\

It is now well established that the simplest approximation to the
electron exchange and correlation (XC) potential, the local-density
approximation (LDA), is inappropriate for treating H bonds.
For example, the dissociation energies of small water clusters and the
cohesive energy of ice are overestimated by $>$50\% with the LDA
\cite{fitzgerald, fitzgerald_hexamer, perinello-ice, Hamann}.
However, despite widespread practical application and several
recent benchmark studies it remains unclear precisely how well the many popular post-LDA
functionals perform at describing H bonds between water clusters.
Generalized gradient approximation (GGA) functionals such as PBE
\cite{PBE}, PW91 \cite{PW91}, and BLYP \cite{Becke88, LYP}, for
example, are widely used to examine liquid water \cite{Todorova,
artacho, tuckerman, parrinello_water,grossman_water}, ice
\cite{Hamann,cerda_angelos_prl,baroni_prl,konig_prl} and adsorbed
water \cite{cerda_angelos_prl,ranea_prl}, yet ask three experts
which one is ``best'' and one is likely to receive three different
answers.
Likewise unanimity has not been reached on the
performance of the many meta-GGA or hybrid functionals that are available,
such as TPSS \cite{TPSS}, PBE0 \cite{PBE0}, and B3LYP
\cite{B3LYP-1,B3LYP-2,B3LYP-3,LYP}.
Part of the reason for the lack of clarity, we believe, stems from
the fact that in previous benchmark studies insufficiently complete
basis sets were employed
and that comparisons were restricted to the simplest H-bonded
systems involving H$_2$O, namely the H$_2$O dimer and trimer.
Basis set incompleteness effects can, of course, mask the true performance
of a given functional and, as we will show below, the ability of a given
functional to accurately predict the strength of the H bond in the dimer or
even the trimer does not necessarily reveal how well that functional will perform
even for the next largest clusters, tetramers and pentamers. \\

In the following we report a study in which the ability of several
GGA, meta-GGA, and hybrid functionals to compute the
energy and structure of H bonds between H$_2$O molecules is
evaluated.
So as to enable the use of large basis sets, which we
demonstrate approach the complete basis set (CBS) limit,  in the
generation of the benchmark data and the DFT data itself, this study
is limited to the four smallest H$_2$O clusters (dimer, trimer,
tetramer, and pentamer).
In addition, this study is restricted to the
established lowest energy conformer of each cluster
\cite{wales_review,saykally_trimer,day_global_search},
which, for orientation purposes,
we show in Fig. 1. For this admittedly small structural data set we
find that, of the functionals tested, the hybrid X3LYP \cite{X3LYP} and PBE0 \cite{PBE0}
functionals offer the best performance.
Among the regular (pure) GGAs mPWLYP \cite{mPW,LYP}
and PBE1W \cite{Truhlar} perform best.
BLYP \cite{Becke88,LYP} and B3LYP \cite{B3LYP-1,B3LYP-2,B3LYP-3,LYP}
predict too weak H bonds and PBE \cite{PBE} and PW91 \cite{PW91}
display rather variable performance with cluster size.
Although MPWB1K \cite{MPWB1K}, PW6B95 \cite{PW6B95}, and B98 \cite{B98}
were previously shown to offer
outstanding performance for water, this is not now the case
when highly accurate basis sets are used. \\
%
%
%
\\
\textbf{II. Reference data - MP2}

For a systematic benchmark study such as this, reliable reference
data is essential.
Experiment is, in principle, one source of this data.
However, experimental dissociation energies are simply not available or
do not come with sufficiently small error bars for all the H$_2$O
clusters we examine here.
Further, with our aim to systematically evaluate the performance of
many DFT XC functionals it becomes impractical to compute all the
small contributions to the experimental dissociation energy that
come on top of the total electronic dissociation energy - an easily
accessible total energy difference - such as
zero point vibrations, relativistic contributions, etc.
The obvious alternative source of reference data are the results
obtained from correlated quantum chemistry methods such as second
order M$\o$ller Plesset perturbation theory (MP2) \cite{MP2} or
coupled-cluster theory \cite{coupled_cluster}.
Indeed such methods have been widely applied to examine H-bonded
systems
\cite{Truhlar,Xantheas-1,Xantheas-2,Xantheas-3,Truhlar_benchmark_2,
Anchor-dimer,Anchor-trimer,Joel,Tsuzuki,K.S.Kim,Svozil,Nielsen,X3LYP-water,wales_review}.
In particular coupled-cluster with single and double excitations
plus a perturbative correction for connected triples (CCSD(T))
produces essentially ``exact'' answers if sufficiently accurate
basis sets are used.
For example, the best CCSD(T) value for the binding energy of the
water dimer is at 217.6$\pm$2 meV \cite{klopper_pccp} in good
agreement with the appropriate experimental number of 216.8$\pm$30
meV \cite{dimr_experiment_2000,dimr_experiment_1979}.
However, since the computational cost of CCSD(T)
formally scales as $N^7$, where $N$ is the number of basis
functions,
the most extravagant use of computational power is required for
CCSD(T) calculations with large basis sets. MP2, on the other hand,
scales as $N^5$ and when compared to CCSD(T) for water dimers and
trimers at the CBS limit, yields binding energies that differ by no
more than 2 meV/H bond \cite{Anchor-dimer,Anchor-trimer}.
In addition, a recent study of water hexamers using CCSD(T) with an aug-cc-pVTZ
basis set revealed that the MP2 and CCSD(T) dissociation
energies of various hexamer structures differ by
$<$3 meV/H$_2$O \cite{hexamer_ccsdt}.
Thus MP2 is a suitable method for
obtaining reference data with an accuracy to within a few meV/H bond.
Such accuracy, which is well beyond so-called chemical accuracy
(1kcal $\approx$ 43 meV), is essential in studies of H-bonded systems. \\

\begin{figure}
    \begin{center}
	  \epsfig{bbllx=23,bblly=210,bburx=583,bbury=628,clip=,
           file=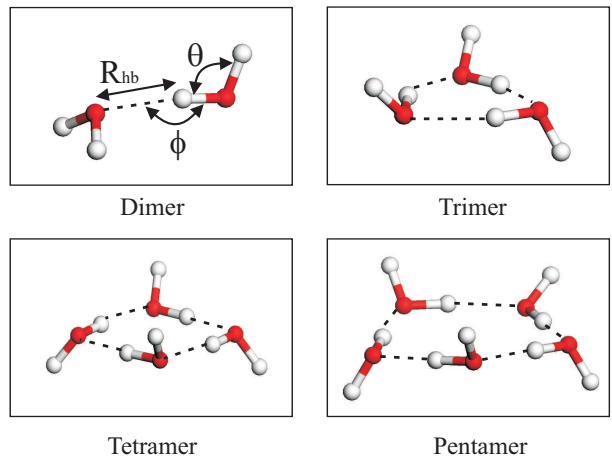,width=8cm}

    \end{center}
    \caption{\label{fig1} Structures of the four water clusters  examined here in their global
    minimum energy configurations. The dashed lines indicate H bonds.
    Some of the structural parameters of the H bond are indicated alongside the
    dimer. We note that in the trimer, tetramer, and pentamer there is one
    H bond per water molecule.}
\end{figure}

Since MP2 geometries are not available for all four clusters
examined here we have computed new MP2 structures for each one. All
calculations have been performed with the Gaussian03 \cite{g03} and
NWChem \cite{nwchem} codes and all geometries were optimized with an
aug-cc-pVTZ basis set within the ``frozen core'' approximation i.e.,
correlations of the oxygen 1$s$ orbital were not considered
\cite{note_g03_nwchem}.
Although the aug-cc-pVTZ basis set is moderately large (92 basis
functions/H$_2$O), this finite basis set will introduce errors in
our predicted MP2 structures.
However, a test with the H$_2$O dimer reveals that the aug-cc-pVTZ
and aug-cc-pVQZ MP2 structures differ by only 0.004~\AA\ in the O-O
bond length and 0.16$^\circ$ in the H bond angle ($\phi$, Fig. 1).
Likewise, Nielsen and co-workers have shown that the MP2 O-O
distances in the cyclic trimer differ by 0.006~\AA\ between the
aug-cc-pVTZ and aug-cc-pVQZ basis sets with all other bonds
differing by $<$0.003~\AA\ \cite{Nielsen}.
For our present purposes these basis set incompleteness errors on
the structures are acceptable and it seems reasonable to assume that
the aug-cc-pVTZ structures reported here come with error bars of
$\pm$0.01~\AA\ for bond lengths and $\pm$0.5$^\circ$ for bond angles.\\

\begin{figure}
    \begin{center}
	  \epsfig{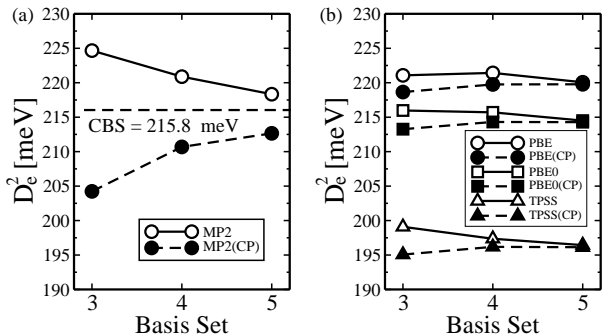}
    \end{center}
    \caption{\label{fig2} (a) Variation in the MP2 dissociation energy for
    the H$_2$O dimer without a counterpoise correction for basis set superposition
    error (BSSE) (labeled MP2) and with a counterpoise correction for BSSE (labeled MP2(CP))
    as a function of basis set size. The extrapolated complete basis set (CBS) dissociation energy for the
    H$_2$O dimer with MP2 is also indicated.
    (b) Variation in the dissociation energy for
    the H$_2$O dimer with and without a counterpoise BSSE correction
    as a function of basis set size
    for three different DFT functionals. The basis set labels
    on the X axis of (a) and (b) indicate aug-cc-pVXZ basis sets,
    where X=3, 4, and 5. Lines are drawn to
    guide the eye only. All structures were optimized with an aug-cc-pVTZ basis set
    consistently with MP2 and with each DFT functional.}
\end{figure}
Total energies and dissociation energies are known to be more
sensitive to basis set incompleteness effects than the geometries
are. To obtain reliable MP2 total energies and dissociation energies
we employ the aug-cc-pVTZ, aug-cc-pVQZ (172 basis functions/H$_2$O)
and aug-cc-pV5Z (287 basis functions/H$_2$O) basis sets in
conjunction with the well-established methods for extrapolating to
the CBS limit. Usually the extrapolation schemes rely on
extrapolating separately the Hartree-Fock (HF) and correlation
contributions to the MP2 total energy. For extrapolation of the HF
part we use Feller's exponential fit \cite{Feller}:
\begin{equation}
     E_X^{HF}=E_{CBS}^{HF}+Ae^{-BX} \quad ,
\label{eqn_HF_extrap}
\end{equation}
where $X$ is the cardinal number corresponding to the basis set
($X$=3, 4, and 5 for the aug-cc-pVTZ, aug-cc-pVQZ, and aug-cc-pV5Z
basis sets, respectively).
%
%
$E_X^{HF}$ is the corresponding HF energy,
$E_{CBS}^{HF}$ is the extrapolated HF energy at the CBS limit, and
$A$ and $B$ are fitting parameters.
For the correlation part of the MP2 total energy we follow an
inverse power of highest angular momentum equation
\cite{Schwartz,Kutzelnigg,Corr-Fit}:
\begin{equation}
E_X^{Corr}=E_{CBS}^{Corr}+CX^{-3}+DX^{-5}\quad ,
\label{eqn_Correlation_extrap}
\end{equation}
where $E_{X}^{Corr}$ is the correlation energy corresponding to $X$,
$E_{CBS}^{Corr}$ is the extrapolated CBS correlation energy, and $C$
and $D$ are fitting parameters.
We have tested various extrapolation schemes available in the
literature \cite{Feller,Schwartz,Kutzelnigg,Corr-Fit,Halkier-1,Truhlar-2,Halkier-2} and did not
see a difference of more than 1.2 meV/H bond between all the predicted CBS
values. We opted for the scheme provided by eqns.
(\ref{eqn_HF_extrap}-\ref{eqn_Correlation_extrap}) because we found
that with input from triple-, quadruple-, and pentuple-$\zeta$ basis sets
this method was best able to predict the total energy of a water monomer
and dimer explicitly calculated with an aug-cc-pV6Z basis set (443 basis
functions/H$_2$O).
%
%
%
%
%
Having obtained MP2 CBS total energies for the H$_2$O monomer and
each of the H$_2$O clusters, we thus arrive at the
MP2 CBS
electronic dissociation energies $(D_{e}^n)$ per H bond which are given by,
\begin{equation}
     D_{e}^n = (E^{nH_2O} - nE^{H_2O})/n_{H-bond} \quad ,
\label{eqn_disso_energy}
\end{equation}
where $E^{nH_2O}$ is the total energy of each cluster with $n$
H$_2$O molecules, $E^{H_2O}$ is the total energy of a H$_2$O
monomer, and $n_{H-bond}$ is the number of H bonds in the cluster.
Our CBS MP2 binding energies for the dimer, trimer, tetramer, and
pentamer are 215.8, 228.5, 299.9, and 314.4 meV/H bond, respectively
\cite{note_HF_CBS}. These values are all within 0.5 meV/H bond of
the previous MP2 CBS dissociation energies reported by Xantheas
\emph{et al.} \cite{Xantheas-2}.
We expect that the various errors accepted in
producing these values (MP2 (valence only) treatment of correlation,
aug-cc-pVTZ structures,
extrapolation to reach the CBS, etc.) will lead to errors in our
reference data from the exact electronic dissociation energies on
the order of $\pm$5.0 meV/H bond at most. With our present aim to
evaluate the performance of various DFT XC functionals such errors
are acceptable.
\\
%
\\
\textbf{III. DFT}

In a study such as this there is an essentially endless list of
functionals that we could consider evaluating. Here we have chosen
to examine  16 different functionals which are widely used or have
been reported to perform particularly well for H-bonded systems in
predicting dissociation energies and structures of the above
mentioned clusters. Specifically we have chosen to optimize
structures of each cluster with the following functionals:
\textbf{(I)} PW91 \cite{PW91} - an extremely popular non-empirical
GGA widely used in calculations of bulk ice
\cite{Hamann,Feibelman,Michaelides_Alavi_King_PRB} and other
H-bonded systems \cite{Tsuzuki};
\textbf{(II)} PBE \cite{PBE} - the twin of PW91 that has again been
widely used and tested for H-bonded systems
\cite{Joel,Truhlar,Truhlar_benchmark_2};
\textbf{(III)} PBE1W - a parameterized empirical variant of PBE specifically
designed to yield improved energetics of H bonds \cite{Truhlar}.
\textbf{(IV)} TPSS \cite{TPSS} - the meta-GGA variant of PBE,
recently used in simulations of liquid water and evaluated for small
water clusters \cite{Truhlar,Perdew-1,Todorova,Truhlar,Truhlar_benchmark_2};
\textbf{(V)} PBE0 \cite{PBE0} - a so-called parameter free hybrid
variant of PBE, also recently tested for water
\cite{Svozil,Todorova,Perdew-2,Truhlar_benchmark_2};
\textbf{(VI)} BLYP - Becke88 \cite{Becke88} exchange combined with LYP
\cite{LYP} correlation, a popular functional for liquid water
simulations \cite{Todorova, artacho, tuckerman, parrinello_water,grossman_water};
\textbf{(VII)} B3LYP \cite{B3LYP-1,B3LYP-2,B3LYP-3,LYP} - the
extremely popular Becke three parameter hybrid functional combined
with LYP nonlocal correlation, which  has, of course, been widely
used to examine H-bonded systems
\cite{Todorova,Svozil,K.S.Kim,Truhlar_benchmark_2};
\textbf{(VIII)} mPWLYP - a combination of a modified PW91 exchange
functional (mPW) \cite{mPW} with the LYP correlation functional,
found to be the most accurate
pure GGA for the energetics of H bonds in water dimers and trimers \cite{Truhlar}.
\textbf{(IX)} BP86 - an empirical GGA combining Becke88
\cite{Becke88} exchange and Perdew86 \cite{Perdew86}
correlation that is well-tested for hydrogen bonded systems
\cite{Perdew-2,Svozil};
\textbf{(X)} X3LYP \cite{X3LYP} - another empirical hybrid functional
designed to describe weak (non-covalent) interactions that
is becoming a familiar name for calculations of water
\cite{X3LYP-water,Todorova,Truhlar_benchmark_2};
\textbf{(XI)} XLYP \cite{X3LYP} - the non-hybrid GGA version of
X3LYP, also tested for H-bonded systems \cite{Truhlar_benchmark_2};
\textbf{(XII)} B98 \cite{B98} - another hybrid functional, said
to perform extremely well for water clusters \cite{Truhlar,Truhlar_benchmark_2};
\textbf{(XIII)} MPWB1K \cite{MPWB1K} - a one parameter hybrid meta-GGA
using mPW \cite{mPW} exchange and Becke95 \cite{B95} correlation,
said to be the joint-best for H bonds between water molecules
\cite{Truhlar,Truhlar_benchmark_2};
\textbf{(XIV)} PW6B95 \cite{PW6B95} - another hybrid meta-GGA
combining PW91 \cite{PW91} exchange and Becke95 \cite{B95}
correlation, found to be the other joint-best functional
for the H bonds between water molecules \cite{Truhlar};
\textbf{(XV)} B3P86 - Becke 3 parameter hybrid functional combined
with Perdew86 nonlocal correlation, found to be best functional for
H-bonded systems in a recent benchmark study
\cite{Truhlar_benchmark_2}; and
\textbf{(XVI)} BH\&HLYP \cite{BH_and_HLYP,Becke88,LYP} - said to
offer similar performance to B3P86
for H-bonded systems \cite{Truhlar_benchmark_2}. \\

\begin{table*}
\caption {\label{tb1} Comparison of the MP2 complete basis set
dissociation energies
to those obtained with various DFT functionals computed with an
aug-cc-pV5Z basis set for four different water clusters. DFT
dissociation energies that come within $\pm$5.0 meV of the
corresponding MP2 value are indicated in bold. The numbers in
parenthesis indicate the percentage cooperative enhancement in the H
bond strength compared to the dissociation energy of the dimer.
Averages of the signed and unsigned errors in the dissociation
energies of all DFT functionals from the corresponding MP2 numbers
over all four clusters are also provided as ME (mean error) and MAE
(mean absolute error). The DFT functionals are ordered in terms of
increasing MAE. All structures were optimized consistently with MP2
and with each DFT functional with an aug-cc-pVTZ basis set and all
values are in meV/H bond (1Kcal/mol = 43.3641 meV).}
\begin{ruledtabular}
\begin{tabular}{c|cccc|cc}
     &  Dimer    &  Trimer        &   Tetramer   &  Pentamer     & ME & MAE \\
\hline
MP2  & 215.8     & 228.5 (5.9)     & 299.9 (38.9) & 314.4 (45.7) & --- & --- \\
X3LYP& \textbf{213.8}& 221.9 (3.8)& \textbf{298.3} (39.5)& \textbf{316.0} (47.8)& -2.2 & 2.9 \\
PBE0 & \textbf{214.5}& \textbf{224.6} (4.7)& \textbf{302.7} (41.1) & 320.9 (49.6) & 1.0 & 3.6 \\
mPWLYP& \textbf{218.5}  & \textbf{226.0} (3.4)& 305.4 (39.8)  & 323.7 (48.1) & 3.8 & 5.0 \\
B3P86 & 203.5    & 220.0 (8.1)     & \textbf{299.4} (47.1)& \textbf{316.5} (55.5) & -4.8 & 5.9  \\
PBE1W& 207.9     & 216.6 (4.0)     & \textbf{294.9} (41.8) & \textbf{312.7} (50.4) & -6.6 & 6.6 \\
BH\&HLYP& \textbf{213.2} & 219.5 (2.9)& 291.3 (36.6)& 308.3 (44.6) & -6.6 & 6.6 \\
PBE  & \textbf{220.1}& \textbf{233.5} (6.1)& 316.4 (43.8) & 334.8 (52.1) & 11.6 & 11.6 \\
B98  & 205.6     & 211.4 (2.8)     & 285.9 (39.1) & 303.1 (47.4)   & -13.2 & 13.2 \\
TPSS & 196.4     & 209.4 (6.6)     & 288.8 (47.0) & 307.5 (56.6)   & -14.1 & 14.1 \\
B3LYP& 197.4     & 206.3 (4.5)     & 280.1 (41.9) & 297.2 (50.6)   & -19.4 & 19.4 \\
PW6B95& 200.9    & 210.5 (4.8)     & 276.8 (37.8) & 292.7 (45.7)   & -19.4 & 19.4 \\
MPWB1K& 199.1    & 210.6 (5.5)     & 276.3 (38.8) & 292.3 (46.8)   & -20.1 & 20.1 \\
BP86 & 184.4     & 205.7 (11.6)    & 282.5 (53.2) & 300.8 (63.1)   & -21.3 & 21.3 \\
PW91 & 232.5     & 244.9 (5.1)     & 330.8 (42.3) & 350.5 (50.8)   & 25.0  & 25.0 \\
XLYP & 191.4     & 198.6 (3.8)     & 272.2 (42.2) & 288.9 (50.9)   & -26.9 & 26.9 \\
BLYP & 180.7     & 191.7 (6.1)     & 264.9 (46.6) & 281.2 (55.6)   & -35.0 & 35.0
\end{tabular}
\end{ruledtabular}
\end{table*}

As with MP2, the question arises as to what basis sets to use in
order to ensure that the DFT results reported here are not
subject
to significant basis set incompleteness errors, which would cloud
our evaluations of the various functionals.
There are no established extrapolation schemes for DFT.
However, it is well-known that DFT total energies are less sensitive
to basis set size than explicitly correlated methods such as MP2
\cite{note_on_cusp,cusp_1_bingel,cusp_2_kato,Helgaker_book}.
Indeed from the plot in Fig. 2 it can be seen that the computed DFT
dissociation energies converge much more rapidly with respect to
basis set size than MP2 does (\emph{c.f.} Figs. 2(a) and (b)).
Specifically we find that upon going from aug-cc-pVTZ to aug-cc-pV5Z
the dissociation energy of the H$_2$O dimer changes by only 1.0,
2.7, and 1.5 meV for the PBE, TPSS, and PBE0 functionals,
respectively.
Further, with the aug-cc-pV5Z basis set we find that the counterpoise
corrected and uncorrected dissociation
energies essentially fall on top of each other, with the largest
difference for the dimers and trimers being 0.45 meV/H bond with the
TPSS functional.
In addition, upon going beyond aug-cc-pV5Z to aug-cc-pV6Z the
dimer dissociation energies change by only 0.24,
0.11, 0.19, 0.25 meV for the PBE, TPSS, PBE0 and BLYP functionals,
respectively.
Thus the DFT dissociation energies we report in the following will all
come from those obtained with the aug-cc-pV5Z basis set, which is
sufficiently large to reflect the true performance
of each functional at a level of accuracy that is reasonably
expected to approach the
basis set limit to within about 0.5 meV/H bond or better.\\
\\
%
%
%
\textbf{IV. Results}
\\
\textbf{A. Dissociation energy}

\begin{figure*}
    \begin{center}
	  \epsfig{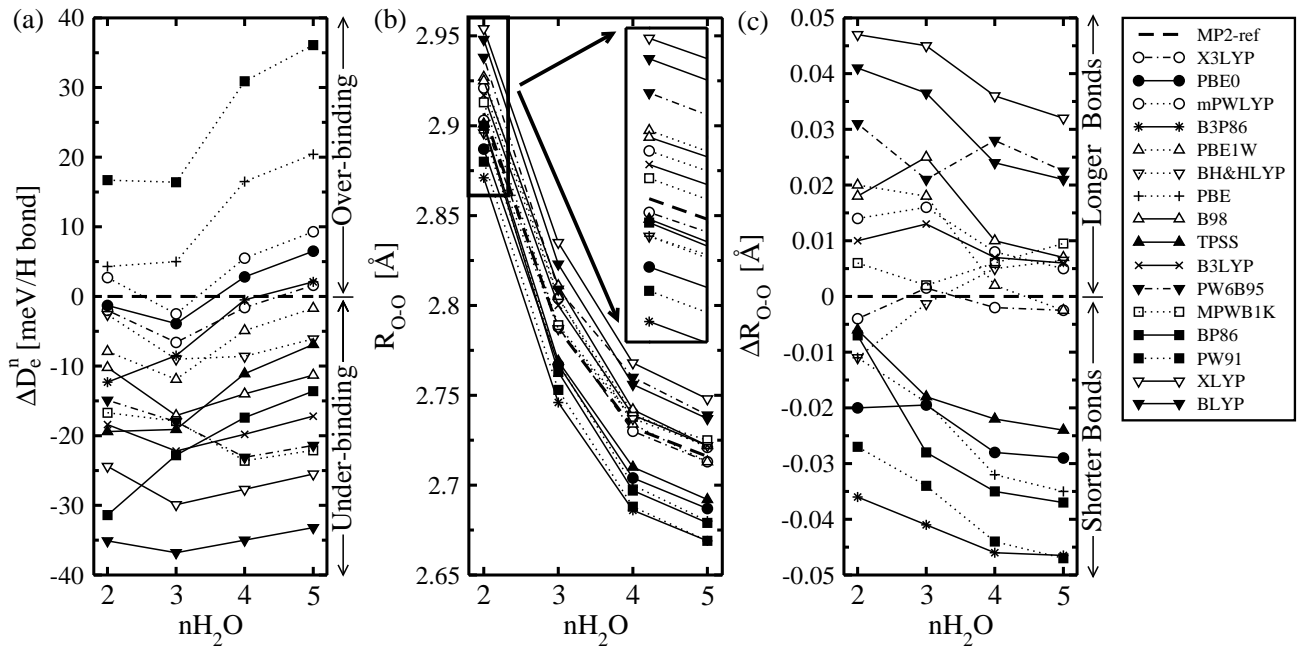}

    \end{center}
    \caption{\label{fig3}
    (a) Difference in the dissociation energy $(\Delta \text{D}^\text{n}_\text{e})$ in meV/H bond of the various
    DFT functionals compared to MP2, plotted as a function of cluster size. Positive values correspond
    to an overestimate of the dissociation energy by a given DFT
    functional.
    (b) Average value of the MP2 and DFT O-O distances (R$_{\text{\text{O-O}}}$) as a function of
    cluster size. The inset zooms in on the dimer region.
    (c) Difference in the average O-O distance $(\Delta \text{R}_{\text{O-O}})$ between MP2 and DFT.
    Positive values correspond
    to an overestimate of the average O-O distances by a given DFT
    functional.
    (a)-(c)
    All DFT energies are calculated with an aug-cc-pV5Z basis set on geometries
    optimized consistently with each functional with an aug-cc-pVTZ basis set.
    Lines are drawn to guide the eye only.}
\end{figure*}

\begin{figure}
    \begin{center}
	  \epsfig{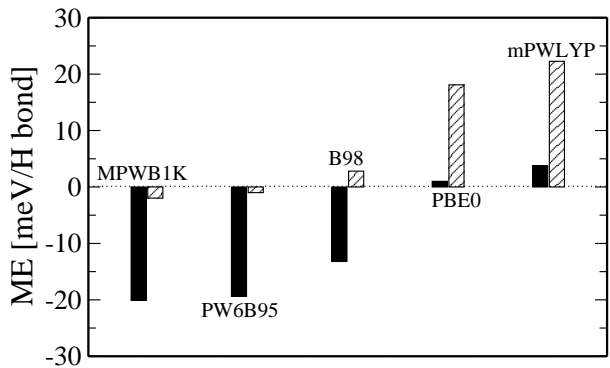}
    \end{center}
    \caption{\label{fig4} Mean error (ME)
    in the dissociation energies obtained with
    aug-cc-pV5Z (solid bars) and MG3S (hashed bars)
    basis sets for five
    selected functionals for the four clusters
    examined here. Positive values correspond to
    an average overestimate of the dissociation energy compared
    to MP2 for the clusters.
    All errors are measured relative to our
    reference CBS MP2 values.}
\end{figure}

In Table \ref{tb1} the computed dissociation energies obtained with MP2
and with each of the DFT functionals are reported.
To allow for a more convenient comparison of the performance of the
various functionals we plot in Fig. \ref{fig3}(a) the difference between
the DFT and MP2 dissociation energies ($\Delta
D_{e}^n$) as a function of water cluster size.
In this figure positive values correspond to an overestimate of the
dissociation energy by a given DFT functional
compared to MP2.
So, what do we learn from Table 1 and  Fig. \ref{fig3}(a)?
First, the functionals which offer the best performance for the
clusters examined are the hybrid X3LYP and PBE0 functionals,
coming within 7 meV/H bond for all four clusters.
Of the non-hybrid functionals the pure GGAs mPWLYP and PBE1W perform best,
coming within 12 meV/H bond for all four clusters.
Second, the very popular BLYP and B3LYP functionals consistently
underbind: B3LYP predicts H bonds
that are $\sim$20 meV too weak; and BLYP predicts H bonds that are
$\sim$35 meV too weak.
%
Third, PBE overestimates the binding in the dimer and trimer ever so
slightly, coming within 5 meV/H bond, but for the tetramer and
pentamer drifts away to yield errors of $\sim$20 meV/H bond.
%
Fourth, PBE and PW91 exhibit a non-negligible difference. Although
it is often assumed that identical numerical results should be
obtained from these two functionals this is not the case here; PW91
is consistently 12-14 meV/H bond worse than PBE. Both functionals,
however, exhibit a similar tendency towards increased overbinding as
the cluster size grows.
Indeed it is clear from  Fig. \ref{fig4}(a) that all PBE-related
functionals (PBE, PW91, PBE1W, TPSS, and PBE0) show this trend,
which in the case of TPSS
means that it gets within $\sim$7 meV/H bond for the pentamer
starting from an error of $\sim$20 meV/H bond for the dimer.
Likewise PBE1W gets closer to the reference value as the cluster size grows.
Finally, despite previous suggestions to the contrary
\cite{Truhlar, Perdew-2,Truhlar_benchmark_2},
none of the other functionals particularly stand out: B98 underbinds by
just over 13 meV/H bond, 
and BP86
exhibits a rather strong variation in performance with cluster size,
ranging from a 30 to 14 meV/H bond error.
B3P86 shows similar behavior to BP86, although the magnitude of the error
is much less and indeed for the tetramer and pentamer B3P86 gives values
close (within 3 meV/H bond) to MP2.
MPWB1K and PW6B95 both underbind by
$\geq$20 meV/H bond.\\ 
\\
\textbf{B. Cooperativity}

An important aspect of the energetics of H bonds is that they tend
to undergo cooperative enhancements,  which for the present systems
implies that the average strengths of the H bonds between the water
molecules increases as the number of H bonds increases. The fact
that the H bonds in water clusters undergo cooperative enhancements
is now well established
\cite{Jeffrey_book,saykally_science_review,Xantheas-4}, as is the
importance of cooperativity in many other types of H-bonded systems
\cite{Jeffrey_book,ranea_prl,Joel-2}. Here we have evaluated the
ability of each functional to correctly capture the computed MP2
cooperative enhancement, defined as the percentage increase in the
average H bond strength compared to that in the H$_2$O dimer.
These numbers are reported in parenthesis in
Table \ref{tb1}. We find that all functionals capture the correct
trend, i.e., the average H bond strength increases upon going from
dimer to pentamer. In addition, most functionals get the absolute
percentage enhancement correct to within 5\%. The notable exceptions
are BP86, B3P86, and TPSS which for the tetramer and pentamer predict
cooperative enhancements that exceed the MP2 values by 10-15\%.\\
\begin{table*}
\caption {\label{tb2} Mean absolute error (MAE) of various DFT
functionals from MP2 for five different structural parameters,
averaged over the four water clusters examined here. The numbers in
bold all have MAE $\leq$0.010 {\AA} for bond lengths and
$\leq$0.50$^\circ$ for bond angles. Mean errors (ME) are given in
parenthesis. All structures were optimized consistently with MP2 and
with each DFT functional with an aug-cc-pVTZ basis set. The order of
the functionals is the same as in Table I.}
\begin{ruledtabular}
\begin{tabular}{c|ccccc}
& $\Delta \text{R}_\text{{O-O}}$ (\AA)& $\Delta \text{R}_\text{{hb}}$ (\AA)& $\Delta \text{R}_\text{{O-H}}$ (\AA)
& $\Delta\phi$  $(^{\circ})$ & $\Delta\theta$  $(^{\circ})$ \\
    \hline
X3LYP& \textbf{0.002} (-0.002)& \textbf{0.003} (-0.003) & \textbf{0.001} (0.000) & \textbf{0.21} (0.21) & 1.04 (1.04) \\
PBE0  & 0.024 (-0.024) & 0.023 (-0.023) & \textbf{0.002} (-0.001)& 0.77 (0.77) & 0.69 (0.69) \\
mPWLYP& 0.012 (0.012) & \textbf{0.008} (-0.004)& 0.012 (0.012)& 0.61 (0.47) & 0.51 (0.51) \\
B3P86 & 0.042 (-0.042)& 0.051 (-0.051) & \textbf{0.003} (0.001)& 1.00 (1.00) & 0.77 (0.77)\\
PBE1W & 0.011 (0.009) & \textbf{0.010} (-0.006) & 0.011 (0.011)& 1.13 (1.13) & \textbf{0.13} (0.13)\\
BH\&HLYP& \textbf{0.006} (-0.003) & 0.015 (0.015) & 0.013 (-0.013)& \textbf{0.48} (-0.17) & 1.52 (1.52) \\
PBE  & 0.024 (-0.024) & 0.046 (-0.046)& 0.012 (0.012)& 1.43 (1.21) & \textbf{0.13} (0.13) \\
B98  & 0.016 (0.016)& 0.015 (0.015)& \textbf{0.001} (-0.001)& 0.52 (0.52) & 0.66 (0.66) \\
TPSS & 0.018 (-0.018)& 0.037 (-0.037)& \textbf{0.010} (0.010)& 1.28 (1.25) & \textbf{0.22} (0.22)\\
B3LYP& \textbf{0.009} (0.009)& \textbf{0.007} (0.007) & \textbf{0.001} (0.001)& \textbf{0.31} (0.31) & 0.93 (0.93)\\
PW6B95& 0.026 (0.026)& 0.029 (0.029) & \textbf{0.006} (-0.006)& \textbf{0.29} (0.24) & 0.81 (0.81)\\
MPWB1K& \textbf{0.006} (0.006)& 0.016 (0.016) & 0.012 (-0.012)& \textbf{0.38} (0.31) & 1.09 (1.09)\\
BP86 & 0.028 (-0.028)& 0.051 (-0.051) & 0.014 (0.014)& 1.58 (1.46) & \textbf{0.11} (0.11) \\
PW91 & 0.038 (-0.038)& 0.038 (-0.038) & 0.012 (0.012)& 1.44 (1.21) & \textbf{0.29} (0.29) \\
XLYP & 0.040 (0.040)& 0.028 (0.028) & 0.011 (0.011)& 0.53 (0.49) & \textbf{0.37} (0.37) \\
BLYP & 0.031 (0.031)& 0.015 (0.015) & \textbf{0.009} (0.009)& 0.69 (0.64) & \textbf{0.37} (0.37)
\end{tabular}
\end{ruledtabular}
\end{table*}
\\
\textbf{C. Geometry}

Let us turn now to an assessment of the quality of the structural
predictions made by each DFT functional. The five key structural
parameters of the H$_2$O clusters that we evaluate are:
(i) The distance between adjacent oxygen atoms involved in a H bond,
$\text{R}_\text{{O-O}}$;
(ii) The length of a H bond, given by the distance between the donor
H and the acceptor O, $\text{R}_{\text{O}\cdots \text{H}}=
\text{R}_{\text{hb}}$ (Fig. 1);
(iii) The H bond angle, $\angle{(\text{O}\cdots \text{H-O})}= \phi$
(Fig. 1);
(iv) The internal O-H bond lengths of each water,
$\text{R}_{\text{O-H}}$; and
(v) The internal H-O-H angle of each water,
$\angle({\text{H-O-H}})=\theta$ (Fig. 1).
In Table II the mean absolute error (MAE) and mean error (ME) between the
MP2 values and those obtained from each functional, averaged over
all four clusters, are listed for each of the above parameters.
This provides an immediate overview for how the functionals
perform.
Summarizing the results of this table, we find that
X3LYP, BH\&HLYP, B3LYP, and MPWB1K
perform the best for O-O distances.
All those functionals yield results that are essentially identical
to MP2, coming within our estimated MP2 bond distance error bar of
0.01 \AA.
B3P86 is the worst functional in terms of O-O distances, with a MAE
of 0.04 \AA.
Largely, these conclusions hold for the related quantity,
R$_\text{{hb}}$,
although now
B3P86, BP86, and PBE perform
worst with MAE values of $\sim$0.05 \AA.
In terms of the H bond angle, $\phi$,
X3LYP, B3LYP, PW6B95, MPWB1K, and BH\&HLYP
are essentially identical to
MP2 coming within our estimated MP2 error bar for angles of
0.5$^\circ$ and again PW91, PBE, and BP86 are the worst being
$\sim$1.5$^\circ$ away from MP2.
For the internal O-H bond lengths
no functional is worse than
$\sim$0.015 \AA\ 
and for the internal
H-O-H angles, $\theta$, all functionals are within
$\sim$1.5$^\circ$ of MP2.\\


One specific aspect of the structures of the small cyclic water
clusters examined here, that is known from experiment and previous
calculations \cite{saykally_science_review, gregory_clary_review}
%
%
is that the average O-O distances between the H$_2$O molecules in the
clusters shorten as the cluster size increases.
This trend is, of course, related to the cooperative
enhancement in H bond strengths discussed already.
As can be seen from the plot of computed O-O distances versus
cluster size in Fig. \ref{fig3}(b) all functionals correctly capture this
effect: the $\sim$0.2 \AA\ shortening in the O-O bond distances upon
going from dimer to pentamer predicted by MP2 is also captured by
every DFT functional.
To look at this issue more closely and specifically to examine how
each functional varies with respect to MP2 we plot in Fig. 3(c) the
difference between the MP2 and DFT O-O distances for the four clusters.
Positive values in Fig. \ref{fig3}(c) indicate that the DFT O-O bonds
are longer than the MP2 ones.
We note that the average MP2 O-O distances for the
dimer, trimer, tetramer and pentamer
are 2.907, 2.787, 2.732, and 2.716 \AA, respectively.
As indicated already in our previous discussion,
X3LYP, B3LYP, BH\&HLYP, and MPWB1K
perform the best at predicting the correct O-O bond length
for each cluster; coming within 0.01 \AA\ of the MP2 values on every
occasion.
Indeed the consistent closeness of the X3LYP O-O distances to the
MP2 ones is remarkable.
PBE0 is a little worse than X3LYP for the O-O distances, predicting
bonds which are consistently about 0.02-0.03 \AA\ too short. Of the
other functionals B3P86 stands out as predicting the shortest O-O
distances (always $\sim$0.04 \AA\ less than MP2) and XLYP and BLYP
predict the longest O-O distances, always at least 0.02 \AA\ longer
than MP2.\\
%
%
%
%
%
\\
\textbf{V. Discussion}

%
%


Here we have shown that of the functionals tested X3LYP and PBE0
offer exceptional performance for the H bonds in small water
clusters in their global minimum energy structures.
However, a previous benchmark study on the ability of most of the
functionals considered here
to describe the energetics of H bonds between water molecules
has arrived at somewhat different conclusions \cite{Truhlar}.
Specifically, a
MAE of 19.5 meV/H bond has been reported for PBE0, worse than the
MAE of 3.6 meV/H bond obtained
here.
In addition, MAEs of 5-7 meV/H bond have been reported with the PW6B95, MPWB1K, and
B98 functionals,
suggesting improved performance for these functionals over what we find here.
In that study the so-called MG3S basis set (identical to 6-311+G(2df,2p)
for H$_2$O) was used.
By comparing the performance of the above-mentioned functionals
with the MG3S and the aug-cc-pV5Z basis sets for the four clusters
under consideration here it appears that the incompleteness of
the MG3S basis set is the main reason for the small discrepancy.
The results, illustrated in the histogram in Fig. \ref{fig4}, reveal
that the dissociation energies obtained with the MG3S basis set are
consistently $\sim$18 meV (0.42 kcal/mol) per H bond larger than
those obtained with the aug-cc-pV5Z basis set.
Thus although PW6B95, MPWB1K, and B98 perform well with the MG3S basis set
(all within $\pm$7 meV/H bond of MP2 for the clusters considered
here) all exhibit a propensity to underbind when the more complete
aug-cc-pV5Z basis set is used.
Conversely, PBE0 and one other functional tested, mPWLYP, which predict
too strong H bonds with the MG3S basis set
(MAEs of 18.1 and 22.3 meV/H bond for the PBE0 and mPWLYP
functionals, respectively, for the clusters examined here) actually
perform very well with the
more complete aug-cc-pV5Z basis set
(MAEs of 3.6 and 5.0 meV/H bond for the PBE0 and mPWLYP
functionals, respectively).
The small and systematic overbinding due to the incompleteness of
the MG3S basis set has also been pointed out by Csonka \emph{et al.} \cite{Perdew-1}. \\

Another interesting aspect of the results of the present study is that the performance of
some functionals differs appreciably  from one cluster to another.
For example, PBE is only $\sim$4-5 meV/H bond away from MP2 for the dimer and trimer but
$>$15 meV/H bond away from MP2 for the tetramer and pentamer.
Conversely, TPSS is
$\sim$20 meV/H bond off MP2 for the dimer but within 7 meV/H bond of MP2 for
the pentamer.
Other functionals which show strong variation in performance with
cluster size are PW91, BP86, and B3P86, and the functional
in the admirable position of showing the least variation,
consistently predicting H bonds that are $\sim$35 meV too weak, is BLYP.
The general conclusion of this analysis, however, is that
it is not necessarily sufficient to use the performance of a
given functional for a single system, such as for example the H$_2$O
dimer, as a guide to how that functional will perform for H bonds
between H$_2$O molecules in general.
Indeed the results reported here indicate that H bond
test sets such as the ``W7''
test set \cite{Truhlar} for water would benefit from the inclusion of
structures other than dimers and trimers.\\

We now ask if the results and conclusions arrived at here are of
general relevance to H$_2$O molecules in other environments and to
other types of H-bonded systems.
Some parallels with DFT simulations of liquid H$_2$O can be seen. It
is generally found, for example, that (when everything else is
equivalent) BLYP liquid H$_2$O is less structured (i.e., the first
peak of the O-O radial distribution function (RDF) has a lower
maximum) than PBE liquid H$_2$O
\cite{Todorova,artacho,tuckerman,parrinello_water,grossman_water};
consistent with the weaker H bonds predicted by BLYP compared to
PBE.
Similarly, the first simulations of liquid H$_2$O with hybrid DFT
functionals (B3LYP, X3LYP, and PBE0) have recently been reported
\cite{Todorova} and the trend in the position of the first peak in
the O-O RDF can be interpreted as being consistent with the current
observations.
Specifically it was found (although the error bars are large because
the simulations were short (5 ps)) that the position of the first
peak in the O-O RDF moves to shorter separation upon going from
B3LYP to X3LYP to PBE0, which is consistent with the small decrease
of the O-O distances (Fig. \ref{fig3}(b)) and increase in H bond
strengths along this series (Table I).
Looking at other H-bonded systems with slightly stronger (for
example, NH$_3\cdots$H$_2$O) 
or slightly weaker H bonds (for example, NH$_3\cdots$NH$_3$) than
those considered here it is known, for example, that PBE generally
overestimates these H bond strengths slightly: PBE overestimates
NH$_3\cdots$H$_2$O by $\sim$30 meV and NH$_3\cdots$NH$_3$ by 6 meV
\cite{Joel}. Likewise, BLYP and B3LYP have been shown to
underestimate a range of H-bonded systems by 20-30 meV/H bond
\cite{Tsuzuki}.
However, the general performance of X3LYP and PBE0 for other H
bonded systems has not been evaluated yet in any great detail with
suitably large basis sets. In light of the present results it will
be interesting to see how well these functionals perform for other
H-bonded systems.
Likewise mPWLYP and PBE1W are not widely used. Since they are
pure GGAs (without any contribution from HF exchange)
they will offer computational savings
compared to X3LYP and PBE0, particularly for condensed phase simulations,
and would thus be interesting to explore further for other H-bonded systems.\\
%
%

Finally, we point out that an interesting conclusion of the present study is the non-negligible
difference between the H bond energies predicted by PBE and PW91;
with PW91 consistently being 12-14 meV/H bond worse than PBE.
A similar discrepancy, although in a rather different area of
application - surface and defect formation energies of metals - has
been identified by Mattsson and co-workers \cite{Mattsson}.
Specifically they found that the PW91 and PBE monovacancy formation energies
of Al differed by $\sim$30-40 meV.
As Mattson and co-workers have done, we caution that it does now not
seem wise to expect identical numerical results from PBE and PW91.\\
\\
%
%
\textbf{VI. Conclusions}

In summary, we have computed MP2 CBS values for the dissociation
energies of small H$_2$O clusters (dimer to pentamer) in their
global minimum energy structures.
This data has been used to evaluate the performance of 16 DFT
functionals.
All DFT energies reported here have been obtained
with an aug-cc-pV5Z basis set, which for DFT is sufficiently large
to enable the true performance of each functional to be assessed,
absent from significant basis set incompleteness errors.
Among the
functionals tested we find that PBE0 and X3LYP perform best for the
energetics of the H bonds considered here; always being within 10 meV/H bond of MP2.
In terms of the structures X3LYP offers outstanding performance,
predicting structures essentially identical to MP2 for all four clusters.
Of the pure GGAs considered mPWLYP and PBE1W perform best.
A small but non-negligible difference in the results
obtained with PBE and PW91 has been identified, with PBE
consistently being 12-14 meV/H bond closer to MP2 than PW91.\\

%
In closing we note that,
although X3LYP and PBE0 predict the most accurate H bond energies,
it is important to remember that \emph{all} functionals considered here do reasonably well.
If, for example, one's definition of ``good'' is so-called chemical
accuracy (1kcal/mol $\approx$ 43 meV/H bond)
then it is clear from Fig.
\ref{fig3}(a) that all functionals achieve chemical accuracy for all
clusters.
The problem is, of course, that for bonds as weak as H bonds,
chemical accuracy is a rather loose criterion since it
amounts to around 20-30\% of the total bond strength.
Future work will involve the investigation of larger H$_2$O clusters in
which the ability of DFT functionals to correctly describe the
ordering of low energy isomeric structures becomes crucial.\\
\\
\textbf{Acknowledgments}

This work is supported by the European Commission through the Early
Stage Researcher Training Network MONET, MEST-CT-2005-020908. See
\emph{www.sljus.lu.se/monet}. A.M's work is supported by a EURYI
award. See \emph{www.esf.org/euryi}. We are grateful to Martin Fuchs
and Joel Ireta for helpful comments
on an earlier version of this manuscript.\\
\\
\textbf{Auxiliary Materials}\\
See EPAPS document No. ---- for a database of the coordinates of the
structures (optimized consistently with MP2 and 16 DFT functionals
with an aug-cc-pVTZ basis set) of all the cluster studied here in
xyz format. The total energies of each clusters obtained from MP2
and the 16 DFT functionals are also provided. This document can be
reached through -----.


\begin{thebibliography}{80}
\expandafter\ifx\csname natexlab\endcsname\relax\def\natexlab#1{#1}\fi
\expandafter\ifx\csname bibnamefont\endcsname\relax
  \def\bibnamefont#1{#1}\fi
\expandafter\ifx\csname bibfnamefont\endcsname\relax
  \def\bibfnamefont#1{#1}\fi
\expandafter\ifx\csname citenamefont\endcsname\relax
  \def\citenamefont#1{#1}\fi
\expandafter\ifx\csname url\endcsname\relax
  \def\url#1{\texttt{#1}}\fi
\expandafter\ifx\csname urlprefix\endcsname\relax\def\urlprefix{URL }\fi
\providecommand{\bibinfo}[2]{#2}
\providecommand{\eprint}[2][]{\url{#2}}

\bibitem[{\citenamefont{Jeffrey}(1997)}]{Jeffrey_book}
\bibinfo{author}{\bibfnamefont{G.~A.} \bibnamefont{Jeffrey}},
  \emph{\bibinfo{title}{An Introduction to Hydrogen Bonding}}
  (\bibinfo{publisher}{Oxford University Press, Inc.}, \bibinfo{address}{New
  York}, \bibinfo{year}{1997}).

\bibitem[{\citenamefont{Liu et~al.}(1996)\citenamefont{Liu, Cruzan, and
  Saykally}}]{saykally_science_review}
\bibinfo{author}{\bibfnamefont{K.}~\bibnamefont{Liu}},
  \bibinfo{author}{\bibfnamefont{J.~D.} \bibnamefont{Cruzan}},
  \bibnamefont{and} \bibinfo{author}{\bibfnamefont{R.~J.}
  \bibnamefont{Saykally}}, \bibinfo{journal}{Science}
  \textbf{\bibinfo{volume}{271}}, \bibinfo{pages}{929} (\bibinfo{year}{1996}).

\bibitem[{\citenamefont{Michaelides and Morgenstern}(in
  press)}]{michaelides_morgenstern}
\bibinfo{author}{\bibfnamefont{A.}~\bibnamefont{Michaelides}} \bibnamefont{and}
  \bibinfo{author}{\bibfnamefont{K.}~\bibnamefont{Morgenstern}},
  \bibinfo{journal}{Nature Mater.}  (\bibinfo{year}{in press}).

\bibitem[{\citenamefont{Todorova et~al.}(2006)\citenamefont{Todorova,
  Seitsonen, Hutter, Kuo, and Mundy}}]{Todorova}
\bibinfo{author}{\bibfnamefont{T.}~\bibnamefont{Todorova}},
  \bibinfo{author}{\bibfnamefont{A.~P.} \bibnamefont{Seitsonen}},
  \bibinfo{author}{\bibfnamefont{J.}~\bibnamefont{Hutter}},
  \bibinfo{author}{\bibfnamefont{I.~W.} \bibnamefont{Kuo}}, \bibnamefont{and}
  \bibinfo{author}{\bibfnamefont{C.~J.} \bibnamefont{Mundy}},
  \bibinfo{journal}{J. Phys. Chem. B} \textbf{\bibinfo{volume}{110}},
  \bibinfo{pages}{3685} (\bibinfo{year}{2006}).

\bibitem[{\citenamefont{Fern\'{a}ndez-Serra and Artacho}(2004)}]{artacho}
\bibinfo{author}{\bibfnamefont{M.~V.} \bibnamefont{Fern\'{a}ndez-Serra}}
  \bibnamefont{and} \bibinfo{author}{\bibfnamefont{E.}~\bibnamefont{Artacho}},
  \bibinfo{journal}{J. Chem. Phys.} \textbf{\bibinfo{volume}{121}},
  \bibinfo{pages}{11136} (\bibinfo{year}{2004}).

\bibitem[{\citenamefont{Lee and Tuckerman}(2006)}]{tuckerman}
\bibinfo{author}{\bibfnamefont{H.}~\bibnamefont{Lee}} \bibnamefont{and}
  \bibinfo{author}{\bibfnamefont{M.~E.} \bibnamefont{Tuckerman}},
  \bibinfo{journal}{J. Phys. Chem. A} \textbf{\bibinfo{volume}{110}},
  \bibinfo{pages}{5549} (\bibinfo{year}{2006}).

\bibitem[{\citenamefont{VandeVondele et~al.}(2005)\citenamefont{VandeVondele,
  Mohamed, Krack, Hutter, Sprik, and Parrinello}}]{parrinello_water}
\bibinfo{author}{\bibfnamefont{J.}~\bibnamefont{VandeVondele}},
  \bibinfo{author}{\bibfnamefont{F.}~\bibnamefont{Mohamed}},
  \bibinfo{author}{\bibfnamefont{M.}~\bibnamefont{Krack}},
  \bibinfo{author}{\bibfnamefont{J.}~\bibnamefont{Hutter}},
  \bibinfo{author}{\bibfnamefont{M.}~\bibnamefont{Sprik}}, \bibnamefont{and}
  \bibinfo{author}{\bibfnamefont{M.}~\bibnamefont{Parrinello}},
  \bibinfo{journal}{J. Chem. Phys.} \textbf{\bibinfo{volume}{122}},
  \bibinfo{pages}{014515} (\bibinfo{year}{2005}).

\bibitem[{\citenamefont{Wernet{ \it et al.}}(2004)}]{nilsson_science}
\bibinfo{author}{\bibfnamefont{Ph.}~\bibnamefont{Wernet{ \it et al.}}},
  \bibinfo{journal}{Science} \textbf{\bibinfo{volume}{304}},
  \bibinfo{pages}{995} (\bibinfo{year}{2004}).

\bibitem[{\citenamefont{Smith et~al.}(2004)\citenamefont{Smith, Cappa, Wilson,
  Messer, Cohen, and Saykally}}]{saykally_science_2004}
\bibinfo{author}{\bibfnamefont{J.~D.} \bibnamefont{Smith}},
  \bibinfo{author}{\bibfnamefont{C.~D.} \bibnamefont{Cappa}},
  \bibinfo{author}{\bibfnamefont{K.~V.} \bibnamefont{Wilson}},
  \bibinfo{author}{\bibfnamefont{B.~M.} \bibnamefont{Messer}},
  \bibinfo{author}{\bibfnamefont{R.~C.} \bibnamefont{Cohen}}, \bibnamefont{and}
  \bibinfo{author}{\bibfnamefont{R.~J.} \bibnamefont{Saykally}},
  \bibinfo{journal}{Science} \textbf{\bibinfo{volume}{306}},
  \bibinfo{pages}{851} (\bibinfo{year}{2004}).

\bibitem[{\citenamefont{Grossman et~al.}(1995)\citenamefont{Grossman,
  Schwegler, Draeger, Gygi, and Galli}}]{grossman_water}
\bibinfo{author}{\bibfnamefont{J.~C.} \bibnamefont{Grossman}},
  \bibinfo{author}{\bibfnamefont{E.}~\bibnamefont{Schwegler}},
  \bibinfo{author}{\bibfnamefont{E.~W.} \bibnamefont{Draeger}},
  \bibinfo{author}{\bibfnamefont{F.}~\bibnamefont{Gygi}}, \bibnamefont{and}
  \bibinfo{author}{\bibfnamefont{G.}~\bibnamefont{Galli}}, \bibinfo{journal}{J.
  Chem. Phys.} \textbf{\bibinfo{volume}{102}}, \bibinfo{pages}{1266}
  (\bibinfo{year}{1995}).

\bibitem[{\citenamefont{Lee et~al.}(1995)\citenamefont{Lee, Chen, and
  Fitzgerald}}]{fitzgerald}
\bibinfo{author}{\bibfnamefont{C.}~\bibnamefont{Lee}},
  \bibinfo{author}{\bibfnamefont{H.}~\bibnamefont{Chen}}, \bibnamefont{and}
  \bibinfo{author}{\bibfnamefont{G.}~\bibnamefont{Fitzgerald}},
  \bibinfo{journal}{J. Chem. Phys.} \textbf{\bibinfo{volume}{102}},
  \bibinfo{pages}{1266} (\bibinfo{year}{1995}).

\bibitem[{\citenamefont{Lee et~al.}(1994)\citenamefont{Lee, Chen, and
  Fitzgerald}}]{fitzgerald_hexamer}
\bibinfo{author}{\bibfnamefont{C.}~\bibnamefont{Lee}},
  \bibinfo{author}{\bibfnamefont{H.}~\bibnamefont{Chen}}, \bibnamefont{and}
  \bibinfo{author}{\bibfnamefont{G.}~\bibnamefont{Fitzgerald}},
  \bibinfo{journal}{J. Chem. Phys.} \textbf{\bibinfo{volume}{101}},
  \bibinfo{pages}{4472} (\bibinfo{year}{1994}).

\bibitem[{\citenamefont{Lee et~al.}(1993)\citenamefont{Lee, Vanderbilt,
  Laasonen, Car, and Parrinello}}]{perinello-ice}
\bibinfo{author}{\bibfnamefont{C.}~\bibnamefont{Lee}},
  \bibinfo{author}{\bibfnamefont{D.}~\bibnamefont{Vanderbilt}},
  \bibinfo{author}{\bibfnamefont{K.}~\bibnamefont{Laasonen}},
  \bibinfo{author}{\bibfnamefont{R.}~\bibnamefont{Car}}, \bibnamefont{and}
  \bibinfo{author}{\bibfnamefont{M.}~\bibnamefont{Parrinello}},
  \bibinfo{journal}{Phys. Rev. B} \textbf{\bibinfo{volume}{47}},
  \bibinfo{pages}{4863} (\bibinfo{year}{1993}).

\bibitem[{\citenamefont{Hamann}(1997)}]{Hamann}
\bibinfo{author}{\bibfnamefont{D.~R.} \bibnamefont{Hamann}},
  \bibinfo{journal}{Phys. Rev. B} \textbf{\bibinfo{volume}{55}},
  \bibinfo{pages}{R10157} (\bibinfo{year}{1997}).

\bibitem[{\citenamefont{Perdew et~al.}(1996)\citenamefont{Perdew, Burke, and
  Ernzerhof}}]{PBE}
\bibinfo{author}{\bibfnamefont{J.~P.} \bibnamefont{Perdew}},
  \bibinfo{author}{\bibfnamefont{K.}~\bibnamefont{Burke}}, \bibnamefont{and}
  \bibinfo{author}{\bibfnamefont{M.}~\bibnamefont{Ernzerhof}},
  \bibinfo{journal}{Phys. Rev. Lett.} \textbf{\bibinfo{volume}{77}},
  \bibinfo{pages}{3865} (\bibinfo{year}{1996}).

\bibitem[{\citenamefont{Perdew}(1991)}]{PW91}
\bibinfo{author}{\bibfnamefont{J.~P.} \bibnamefont{Perdew}},
  \emph{\bibinfo{title}{in Electronic Structure of Solids '91}}
  (\bibinfo{publisher}{edited by P. Ziesche and H. Eschrig},
  \bibinfo{address}{Akademie Verlag, Berlin}, \bibinfo{year}{1991}),
  \bibinfo{note}{p. 11}.

\bibitem[{\citenamefont{Becke}(1988)}]{Becke88}
\bibinfo{author}{\bibfnamefont{A.~D.} \bibnamefont{Becke}},
  \bibinfo{journal}{Phys. Rev. A} \textbf{\bibinfo{volume}{38}},
  \bibinfo{pages}{3098} (\bibinfo{year}{1988}).

\bibitem[{\citenamefont{Lee and Parr}(1988)}]{LYP}
\bibinfo{author}{\bibfnamefont{W.}~\bibnamefont{Lee}, \bibfnamefont{C.~Yang}}
  \bibnamefont{and} \bibinfo{author}{\bibfnamefont{R.~G.} \bibnamefont{Parr}},
  \bibinfo{journal}{Phys. Rev. B} \textbf{\bibinfo{volume}{37}},
  \bibinfo{pages}{785} (\bibinfo{year}{1988}).

\bibitem[{\citenamefont{Cerd\'{a}{ \it et al.}}(2004)}]{cerda_angelos_prl}
\bibinfo{author}{\bibfnamefont{J.}~\bibnamefont{Cerd\'{a}{ \it et al.}}},
  \bibinfo{journal}{Phys. Rev. Lett.} \textbf{\bibinfo{volume}{93}},
  \bibinfo{pages}{116101} (\bibinfo{year}{2004}).

\bibitem[{\citenamefont{Umemoto et~al.}(2004)\citenamefont{Umemoto,
  Wentzcovitch, Baroni, and Gironcoli}}]{baroni_prl}
\bibinfo{author}{\bibfnamefont{K.}~\bibnamefont{Umemoto}},
  \bibinfo{author}{\bibfnamefont{R.~M.} \bibnamefont{Wentzcovitch}},
  \bibinfo{author}{\bibfnamefont{S.}~\bibnamefont{Baroni}}, \bibnamefont{and}
  \bibinfo{author}{\bibfnamefont{S.}~\bibnamefont{Gironcoli}},
  \bibinfo{journal}{Phys. Rev. Lett.} \textbf{\bibinfo{volume}{92}},
  \bibinfo{pages}{105502} (\bibinfo{year}{2004}).

\bibitem[{\citenamefont{de~Koning et~al.}(2006)\citenamefont{de~Koning,
  Antonelli, da~Silva, and Fazzio}}]{konig_prl}
\bibinfo{author}{\bibfnamefont{M.}~\bibnamefont{de~Koning}},
  \bibinfo{author}{\bibfnamefont{A.}~\bibnamefont{Antonelli}},
  \bibinfo{author}{\bibfnamefont{A.~J.~R.} \bibnamefont{da~Silva}},
  \bibnamefont{and} \bibinfo{author}{\bibfnamefont{A.}~\bibnamefont{Fazzio}},
  \bibinfo{journal}{Phys. Rev. Lett.} \textbf{\bibinfo{volume}{97}},
  \bibinfo{pages}{155501} (\bibinfo{year}{2006}).

\bibitem[{\citenamefont{Ranea{ \it et al.}}(2004)}]{ranea_prl}
\bibinfo{author}{\bibfnamefont{V.~A.} \bibnamefont{Ranea{ \it et al.}}},
  \bibinfo{journal}{Phys. Rev. Lett.} \textbf{\bibinfo{volume}{92}},
  \bibinfo{pages}{136104} (\bibinfo{year}{2004}).

\bibitem[{\citenamefont{Tao et~al.}(2003)\citenamefont{Tao, Perdew, Staroverov,
  and Scuseria}}]{TPSS}
\bibinfo{author}{\bibfnamefont{J.}~\bibnamefont{Tao}},
  \bibinfo{author}{\bibfnamefont{J.~P.} \bibnamefont{Perdew}},
  \bibinfo{author}{\bibfnamefont{V.~N.} \bibnamefont{Staroverov}},
  \bibnamefont{and} \bibinfo{author}{\bibfnamefont{G.~E.}
  \bibnamefont{Scuseria}}, \bibinfo{journal}{Phys. Rev. Lett.}
  \textbf{\bibinfo{volume}{91}}, \bibinfo{pages}{146401}
  (\bibinfo{year}{2003}).

\bibitem[{\citenamefont{Adamo and Barone}(1999)}]{PBE0}
\bibinfo{author}{\bibfnamefont{C.}~\bibnamefont{Adamo}} \bibnamefont{and}
  \bibinfo{author}{\bibfnamefont{V.}~\bibnamefont{Barone}},
  \bibinfo{journal}{J. Chem. Phys.} \textbf{\bibinfo{volume}{110}},
  \bibinfo{pages}{6158} (\bibinfo{year}{1999}).

\bibitem[{\citenamefont{Becke}(1993{\natexlab{a}})}]{B3LYP-1}
\bibinfo{author}{\bibfnamefont{A.~D.} \bibnamefont{Becke}},
  \bibinfo{journal}{J. Chem. Phys.} \textbf{\bibinfo{volume}{98}},
  \bibinfo{pages}{5648} (\bibinfo{year}{1993}{\natexlab{a}}).

\bibitem[{\citenamefont{Vosko et~al.}(1980)\citenamefont{Vosko, Wilk, and
  Nusair}}]{B3LYP-2}
\bibinfo{author}{\bibfnamefont{S.~H.} \bibnamefont{Vosko}},
  \bibinfo{author}{\bibfnamefont{L.}~\bibnamefont{Wilk}}, \bibnamefont{and}
  \bibinfo{author}{\bibfnamefont{M.}~\bibnamefont{Nusair}},
  \bibinfo{journal}{Can. J. Phys.} \textbf{\bibinfo{volume}{58}},
  \bibinfo{pages}{1200} (\bibinfo{year}{1980}).

\bibitem[{\citenamefont{Stephens et~al.}(1994)\citenamefont{Stephens, Devlin,
  Chabalowski, and Frisch}}]{B3LYP-3}
\bibinfo{author}{\bibfnamefont{P.~J.} \bibnamefont{Stephens}},
  \bibinfo{author}{\bibfnamefont{F.~J.} \bibnamefont{Devlin}},
  \bibinfo{author}{\bibfnamefont{C.~F.} \bibnamefont{Chabalowski}},
  \bibnamefont{and} \bibinfo{author}{\bibfnamefont{M.~J.}
  \bibnamefont{Frisch}}, \bibinfo{journal}{J. Phys. Chem.}
  \textbf{\bibinfo{volume}{98}}, \bibinfo{pages}{11623} (\bibinfo{year}{1994}).

\bibitem[{\citenamefont{Taketsugu and Wales}(2002)}]{wales_review}
\bibinfo{author}{\bibfnamefont{T.}~\bibnamefont{Taketsugu}} \bibnamefont{and}
  \bibinfo{author}{\bibfnamefont{D.~J.} \bibnamefont{Wales}},
  \bibinfo{journal}{Mol. Phys.} \textbf{\bibinfo{volume}{100}},
  \bibinfo{pages}{2793} (\bibinfo{year}{2002}).

\bibitem[{\citenamefont{Keutsch et~al.}(2003)\citenamefont{Keutsch, Cruzan, and
  Saykally}}]{saykally_trimer}
\bibinfo{author}{\bibfnamefont{F.~N.} \bibnamefont{Keutsch}},
  \bibinfo{author}{\bibfnamefont{J.~D.} \bibnamefont{Cruzan}},
  \bibnamefont{and} \bibinfo{author}{\bibfnamefont{R.~J.}
  \bibnamefont{Saykally}}, \bibinfo{journal}{Chem. Rev.}
  \textbf{\bibinfo{volume}{103}}, \bibinfo{pages}{2533} (\bibinfo{year}{2003}).

\bibitem[{\citenamefont{Day et~al.}(2005)\citenamefont{Day, Kirschner, and
  Shields}}]{day_global_search}
\bibinfo{author}{\bibfnamefont{M.~B.} \bibnamefont{Day}},
  \bibinfo{author}{\bibfnamefont{K.~N.} \bibnamefont{Kirschner}},
  \bibnamefont{and} \bibinfo{author}{\bibfnamefont{G.~C.}
  \bibnamefont{Shields}}, \bibinfo{journal}{J. Phys. Chem. A}
  \textbf{\bibinfo{volume}{109}}, \bibinfo{pages}{6773} (\bibinfo{year}{2005}).

\bibitem[{\citenamefont{Xu and Goddard~III}(2004)}]{X3LYP}
\bibinfo{author}{\bibfnamefont{X.}~\bibnamefont{Xu}} \bibnamefont{and}
  \bibinfo{author}{\bibfnamefont{W.~A.} \bibnamefont{Goddard~III}},
  \bibinfo{journal}{Proc. Natl. Acad. Sci. U.S.A.}
  \textbf{\bibinfo{volume}{101}}, \bibinfo{pages}{2673} (\bibinfo{year}{2004}).

\bibitem[{\citenamefont{Adamo and Barone}(1998)}]{mPW}
\bibinfo{author}{\bibfnamefont{C.}~\bibnamefont{Adamo}} \bibnamefont{and}
  \bibinfo{author}{\bibfnamefont{V.}~\bibnamefont{Barone}},
  \bibinfo{journal}{J. Chem. Phys.} \textbf{\bibinfo{volume}{108}},
  \bibinfo{pages}{664} (\bibinfo{year}{1998}).

\bibitem[{\citenamefont{Dahlke and Truhlar}(2005)}]{Truhlar}
\bibinfo{author}{\bibfnamefont{E.~E.} \bibnamefont{Dahlke}} \bibnamefont{and}
  \bibinfo{author}{\bibfnamefont{D.~G.} \bibnamefont{Truhlar}},
  \bibinfo{journal}{J. Phys. Chem. B} \textbf{\bibinfo{volume}{109}},
  \bibinfo{pages}{15677} (\bibinfo{year}{2005}).

\bibitem[{\citenamefont{Zhao and Truhlar}(2004)}]{MPWB1K}
\bibinfo{author}{\bibfnamefont{Y.}~\bibnamefont{Zhao}} \bibnamefont{and}
  \bibinfo{author}{\bibfnamefont{D.~G.} \bibnamefont{Truhlar}},
  \bibinfo{journal}{J. Phys. Chem. A} \textbf{\bibinfo{volume}{108}},
  \bibinfo{pages}{6908} (\bibinfo{year}{2004}).

\bibitem[{\citenamefont{Zhao and Truhlar}(2005{\natexlab{a}})}]{PW6B95}
\bibinfo{author}{\bibfnamefont{Y.}~\bibnamefont{Zhao}} \bibnamefont{and}
  \bibinfo{author}{\bibfnamefont{D.~G.} \bibnamefont{Truhlar}},
  \bibinfo{journal}{J. Phys. Chem. A} \textbf{\bibinfo{volume}{109}},
  \bibinfo{pages}{5656} (\bibinfo{year}{2005}{\natexlab{a}}).

\bibitem[{\citenamefont{Schmider and Becke}(1998)}]{B98}
\bibinfo{author}{\bibfnamefont{H.~L.} \bibnamefont{Schmider}} \bibnamefont{and}
  \bibinfo{author}{\bibfnamefont{A.~D.} \bibnamefont{Becke}},
  \bibinfo{journal}{J. Chem. Phys.} \textbf{\bibinfo{volume}{108}},
  \bibinfo{pages}{9624} (\bibinfo{year}{1998}).

\bibitem[{\citenamefont{M$\o$ller and Plesset}(1934)}]{MP2}
\bibinfo{author}{\bibfnamefont{C.}~\bibnamefont{M$\o$ller}} \bibnamefont{and}
  \bibinfo{author}{\bibfnamefont{M.~S.} \bibnamefont{Plesset}},
  \bibinfo{journal}{Phys. Rev.} \textbf{\bibinfo{volume}{46}},
  \bibinfo{pages}{618} (\bibinfo{year}{1934}).

\bibitem[{\citenamefont{Coester and K\"{u}mmel}(1960)}]{coupled_cluster}
\bibinfo{author}{\bibfnamefont{F.}~\bibnamefont{Coester}} \bibnamefont{and}
  \bibinfo{author}{\bibfnamefont{H.}~\bibnamefont{K\"{u}mmel}},
  \bibinfo{journal}{Nucl. Phys.} \textbf{\bibinfo{volume}{17}},
  \bibinfo{pages}{477} (\bibinfo{year}{1960}).

\bibitem[{\citenamefont{Xantheas}(1994)}]{Xantheas-1}
\bibinfo{author}{\bibfnamefont{S.~S.} \bibnamefont{Xantheas}},
  \bibinfo{journal}{J. Chem. Phys.} \textbf{\bibinfo{volume}{102}},
  \bibinfo{pages}{4505} (\bibinfo{year}{1994}).

\bibitem[{\citenamefont{Xantheas et~al.}(2002)\citenamefont{Xantheas, Burnham,
  and Harison}}]{Xantheas-2}
\bibinfo{author}{\bibfnamefont{S.~S.} \bibnamefont{Xantheas}},
  \bibinfo{author}{\bibfnamefont{C.~J.} \bibnamefont{Burnham}},
  \bibnamefont{and} \bibinfo{author}{\bibfnamefont{R.~J.}
  \bibnamefont{Harison}}, \bibinfo{journal}{J. Chem. Phys.}
  \textbf{\bibinfo{volume}{116}}, \bibinfo{pages}{1493} (\bibinfo{year}{2002}).

\bibitem[{\citenamefont{Xantheas and Apr\`{a}}(2004)}]{Xantheas-3}
\bibinfo{author}{\bibfnamefont{S.~S.} \bibnamefont{Xantheas}} \bibnamefont{and}
  \bibinfo{author}{\bibfnamefont{E.}~\bibnamefont{Apr\`{a}}},
  \bibinfo{journal}{J. Chem. Phys.} \textbf{\bibinfo{volume}{120}},
  \bibinfo{pages}{823} (\bibinfo{year}{2004}).

\bibitem[{\citenamefont{Zhao and
  Truhlar}(2005{\natexlab{b}})}]{Truhlar_benchmark_2}
\bibinfo{author}{\bibfnamefont{Y.}~\bibnamefont{Zhao}} \bibnamefont{and}
  \bibinfo{author}{\bibfnamefont{D.~G.} \bibnamefont{Truhlar}},
  \bibinfo{journal}{J. Chem. Theory Comput.} \textbf{\bibinfo{volume}{1}},
  \bibinfo{pages}{415} (\bibinfo{year}{2005}{\natexlab{b}}).

\bibitem[{\citenamefont{Tschumper et~al.}(2002)\citenamefont{Tschumper,
  Leininger, Hoffman, Valeev, Schaefer~III, and Quack}}]{Anchor-dimer}
\bibinfo{author}{\bibfnamefont{G.~S.} \bibnamefont{Tschumper}},
  \bibinfo{author}{\bibfnamefont{M.~L.} \bibnamefont{Leininger}},
  \bibinfo{author}{\bibfnamefont{B.~C.} \bibnamefont{Hoffman}},
  \bibinfo{author}{\bibfnamefont{E.~F.} \bibnamefont{Valeev}},
  \bibinfo{author}{\bibfnamefont{H.~F.} \bibnamefont{Schaefer~III}},
  \bibnamefont{and} \bibinfo{author}{\bibfnamefont{M.}~\bibnamefont{Quack}},
  \bibinfo{journal}{J. Chem. Phys.} \textbf{\bibinfo{volume}{116}},
  \bibinfo{pages}{690} (\bibinfo{year}{2002}).

\bibitem[{\citenamefont{Anderson et~al.}(2004)\citenamefont{Anderson, Crager,
  Fedoroff, and Tschumper}}]{Anchor-trimer}
\bibinfo{author}{\bibfnamefont{J.~A.} \bibnamefont{Anderson}},
  \bibinfo{author}{\bibfnamefont{K.}~\bibnamefont{Crager}},
  \bibinfo{author}{\bibfnamefont{L.}~\bibnamefont{Fedoroff}}, \bibnamefont{and}
  \bibinfo{author}{\bibfnamefont{G.~S.} \bibnamefont{Tschumper}},
  \bibinfo{journal}{J. Chem. Phys.} \textbf{\bibinfo{volume}{121}},
  \bibinfo{pages}{11023} (\bibinfo{year}{2004}).

\bibitem[{\citenamefont{Ireta et~al.}(2004)\citenamefont{Ireta, Neugebauer, and
  Scheffler}}]{Joel}
\bibinfo{author}{\bibfnamefont{J.}~\bibnamefont{Ireta}},
  \bibinfo{author}{\bibfnamefont{J.}~\bibnamefont{Neugebauer}},
  \bibnamefont{and}
  \bibinfo{author}{\bibfnamefont{M.}~\bibnamefont{Scheffler}},
  \bibinfo{journal}{J. Phys. Chem. A} \textbf{\bibinfo{volume}{108}},
  \bibinfo{pages}{5692} (\bibinfo{year}{2004}).

\bibitem[{\citenamefont{Tsuzuki and Luthi}(2001)}]{Tsuzuki}
\bibinfo{author}{\bibfnamefont{T.}~\bibnamefont{Tsuzuki}} \bibnamefont{and}
  \bibinfo{author}{\bibfnamefont{H.~P.} \bibnamefont{Luthi}},
  \bibinfo{journal}{J. Chem. Phys.} \textbf{\bibinfo{volume}{114}},
  \bibinfo{pages}{3949} (\bibinfo{year}{2001}).

\bibitem[{\citenamefont{Lee et~al.}(2000)\citenamefont{Lee, Suh, Lee,
  Tarakeshwar, and Kim}}]{K.S.Kim}
\bibinfo{author}{\bibfnamefont{H.~M.} \bibnamefont{Lee}},
  \bibinfo{author}{\bibfnamefont{S.~B.} \bibnamefont{Suh}},
  \bibinfo{author}{\bibfnamefont{J.~Y.} \bibnamefont{Lee}},
  \bibinfo{author}{\bibfnamefont{P.}~\bibnamefont{Tarakeshwar}},
  \bibnamefont{and} \bibinfo{author}{\bibfnamefont{K.~S.} \bibnamefont{Kim}},
  \bibinfo{journal}{J. Chem. Phys.} \textbf{\bibinfo{volume}{112}},
  \bibinfo{pages}{9759} (\bibinfo{year}{2000}).

\bibitem[{\citenamefont{Svozil and Jungwirth}(2006)}]{Svozil}
\bibinfo{author}{\bibfnamefont{D.}~\bibnamefont{Svozil}} \bibnamefont{and}
  \bibinfo{author}{\bibfnamefont{P.}~\bibnamefont{Jungwirth}},
  \bibinfo{journal}{J. Phys. Chem. A} \textbf{\bibinfo{volume}{110}},
  \bibinfo{pages}{9194} (\bibinfo{year}{2006}).

\bibitem[{\citenamefont{Nielsen et~al.}(1999)\citenamefont{Nielsen, Seidl, and
  Janssen}}]{Nielsen}
\bibinfo{author}{\bibfnamefont{I.~M.~B.} \bibnamefont{Nielsen}},
  \bibinfo{author}{\bibfnamefont{E.~T.} \bibnamefont{Seidl}}, \bibnamefont{and}
  \bibinfo{author}{\bibfnamefont{C.~L.} \bibnamefont{Janssen}},
  \bibinfo{journal}{J. Chem. Phys.} \textbf{\bibinfo{volume}{110}},
  \bibinfo{pages}{9435} (\bibinfo{year}{1999}).

\bibitem[{\citenamefont{Su et~al.}(2004)\citenamefont{Su, Xu, and
  Goddard~III}}]{X3LYP-water}
\bibinfo{author}{\bibfnamefont{J.~T.} \bibnamefont{Su}},
  \bibinfo{author}{\bibfnamefont{X.}~\bibnamefont{Xu}}, \bibnamefont{and}
  \bibinfo{author}{\bibfnamefont{W.~A.} \bibnamefont{Goddard~III}},
  \bibinfo{journal}{J. Phys. Chem. A} \textbf{\bibinfo{volume}{108}},
  \bibinfo{pages}{10518} (\bibinfo{year}{2004}).

\bibitem[{\citenamefont{Klopper et~al.}(2000)\citenamefont{Klopper, van
  Duijneveldt-van~de Rijdt, and van Duijneveldt}}]{klopper_pccp}
\bibinfo{author}{\bibfnamefont{W.}~\bibnamefont{Klopper}},
  \bibinfo{author}{\bibfnamefont{J.~G. C.~M.} \bibnamefont{van
  Duijneveldt-van~de Rijdt}}, \bibnamefont{and}
  \bibinfo{author}{\bibfnamefont{F.~B.} \bibnamefont{van Duijneveldt}},
  \bibinfo{journal}{Phys. Chem. Chem. Phys.} \textbf{\bibinfo{volume}{2}},
  \bibinfo{pages}{2227} (\bibinfo{year}{2000}).

\bibitem[{\citenamefont{Mas et~al.}(2000)\citenamefont{Mas, Bukowski,
  Szalewicz, Groenenboom, Wormer, and van~der Avoird}}]{dimr_experiment_2000}
\bibinfo{author}{\bibfnamefont{E.~M.} \bibnamefont{Mas}},
  \bibinfo{author}{\bibfnamefont{R.}~\bibnamefont{Bukowski}},
  \bibinfo{author}{\bibfnamefont{K.}~\bibnamefont{Szalewicz}},
  \bibinfo{author}{\bibfnamefont{G.~C.} \bibnamefont{Groenenboom}},
  \bibinfo{author}{\bibfnamefont{P.~E.~S.} \bibnamefont{Wormer}},
  \bibnamefont{and} \bibinfo{author}{\bibfnamefont{A.}~\bibnamefont{van~der
  Avoird}}, \bibinfo{journal}{J. Chem. Phys.} \textbf{\bibinfo{volume}{113}},
  \bibinfo{pages}{6687} (\bibinfo{year}{2000}).

\bibitem[{\citenamefont{Curtiss et~al.}(1979)\citenamefont{Curtiss, Frurip, and
  Blander}}]{dimr_experiment_1979}
\bibinfo{author}{\bibfnamefont{L.~A.} \bibnamefont{Curtiss}},
  \bibinfo{author}{\bibfnamefont{D.~J.} \bibnamefont{Frurip}},
  \bibnamefont{and} \bibinfo{author}{\bibfnamefont{M.}~\bibnamefont{Blander}},
  \bibinfo{journal}{J. Chem. Phys.} \textbf{\bibinfo{volume}{71}},
  \bibinfo{pages}{2703} (\bibinfo{year}{1979}).

\bibitem[{\citenamefont{Olson et~al.}(in press)\citenamefont{Olson, Bentz,
  Kendall, Schmidt, and Gordon}}]{hexamer_ccsdt}
\bibinfo{author}{\bibfnamefont{R.~M.} \bibnamefont{Olson}},
  \bibinfo{author}{\bibfnamefont{J.~L.} \bibnamefont{Bentz}},
  \bibinfo{author}{\bibfnamefont{R.~A.} \bibnamefont{Kendall}},
  \bibinfo{author}{\bibfnamefont{M.~W.} \bibnamefont{Schmidt}},
  \bibnamefont{and} \bibinfo{author}{\bibfnamefont{M.~S.}
  \bibnamefont{Gordon}}, \bibinfo{journal}{J. Chem. Theory Comput.}
  (\bibinfo{year}{in press}).

\bibitem[{\citenamefont{Frisch{ \it et al.}}(2004)}]{g03}
\bibinfo{author}{\bibfnamefont{M.~J.} \bibnamefont{Frisch{ \it et al.}}},
  \bibinfo{journal}{Gaussian 03. Revision C.02; Gaussian, Inc.}
  (\bibinfo{year}{2004}).

\bibitem[{\citenamefont{Bylaska{ \it et al.}}(2006)}]{nwchem}
\bibinfo{author}{\bibfnamefont{E.~J.} \bibnamefont{Bylaska{ \it et al.}}},
  \bibinfo{journal}{`NWChem, A Computational Chemistry Package for Parallel
  Computers, Version 5.0', Pacific Northwest National Laboratory, Richland,
  Washington 99352-0999, USA}  (\bibinfo{year}{2006}).

\bibitem[{not({\natexlab{a}})}]{note_g03_nwchem}
\bibinfo{note}{We have used Gaussian03 \cite{g03} and NWChem \cite{nwchem}
  interchangeably, since total energies of the water clusters obtained from the
  two codes differ by no more than 0.4 meV per water molecule.}

\bibitem[{\citenamefont{Feller}(1993)}]{Feller}
\bibinfo{author}{\bibfnamefont{D.}~\bibnamefont{Feller}}, \bibinfo{journal}{J.
  Chem. Phys.} \textbf{\bibinfo{volume}{98}}, \bibinfo{pages}{7059}
  (\bibinfo{year}{1993}).

\bibitem[{\citenamefont{Schwartz}(1962)}]{Schwartz}
\bibinfo{author}{\bibfnamefont{C.}~\bibnamefont{Schwartz}},
  \bibinfo{journal}{Phys. Rev.} \textbf{\bibinfo{volume}{126}},
  \bibinfo{pages}{1015} (\bibinfo{year}{1962}).

\bibitem[{\citenamefont{Kutzelnigg and Morgan~III}(1992)}]{Kutzelnigg}
\bibinfo{author}{\bibfnamefont{W.}~\bibnamefont{Kutzelnigg}} \bibnamefont{and}
  \bibinfo{author}{\bibfnamefont{J.~D.} \bibnamefont{Morgan~III}},
  \bibinfo{journal}{J. Chem. Phys.} \textbf{\bibinfo{volume}{96}},
  \bibinfo{pages}{4484} (\bibinfo{year}{1992}).

\bibitem[{\citenamefont{Willson and Dunning~Jr.}(1997)}]{Corr-Fit}
\bibinfo{author}{\bibfnamefont{A.~K.} \bibnamefont{Willson}} \bibnamefont{and}
  \bibinfo{author}{\bibfnamefont{T.~H.} \bibnamefont{Dunning~Jr.}},
  \bibinfo{journal}{J. Chem. Phys.} \textbf{\bibinfo{volume}{106}},
  \bibinfo{pages}{8718} (\bibinfo{year}{1997}).

\bibitem[{\citenamefont{Halkier{ \it et al.}}(1998)}]{Halkier-1}
\bibinfo{author}{\bibfnamefont{A.}~\bibnamefont{Halkier{ \it et al.}}},
  \bibinfo{journal}{Chem. Phys. Lett.} \textbf{\bibinfo{volume}{286}},
  \bibinfo{pages}{243} (\bibinfo{year}{1998}).

\bibitem[{\citenamefont{Truhlar}(1998)}]{Truhlar-2}
\bibinfo{author}{\bibfnamefont{D.~G.} \bibnamefont{Truhlar}},
  \bibinfo{journal}{Chem. Phys. Lett.} \textbf{\bibinfo{volume}{294}},
  \bibinfo{pages}{45} (\bibinfo{year}{1998}).

\bibitem[{\citenamefont{Halkier{ \it et al.}}(1999)}]{Halkier-2}
\bibinfo{author}{\bibfnamefont{A.}~\bibnamefont{Halkier{ \it et al.}}},
  \bibinfo{journal}{J. Chem. Phys.} \textbf{\bibinfo{volume}{111}},
  \bibinfo{pages}{9157} (\bibinfo{year}{1999}).

\bibitem[{not({\natexlab{b}})}]{note_HF_CBS}
\bibinfo{note}{The corresponding CBS HF dissociation energies at the MP2
  structures are 148.0, 140.8, 187.6, and 200.8 meV/H bond for the dimer,
  trimer, tetramer, and pentamer, respectively.}

\bibitem[{\citenamefont{Feibelman}(2002)}]{Feibelman}
\bibinfo{author}{\bibfnamefont{P.~J.} \bibnamefont{Feibelman}},
  \bibinfo{journal}{Science} \textbf{\bibinfo{volume}{295}},
  \bibinfo{pages}{99} (\bibinfo{year}{2002}).

\bibitem[{\citenamefont{Michaelides et~al.}(2004)\citenamefont{Michaelides,
  Alavi, and King}}]{Michaelides_Alavi_King_PRB}
\bibinfo{author}{\bibfnamefont{A.}~\bibnamefont{Michaelides}},
  \bibinfo{author}{\bibfnamefont{A.}~\bibnamefont{Alavi}}, \bibnamefont{and}
  \bibinfo{author}{\bibfnamefont{D.~A.} \bibnamefont{King}},
  \bibinfo{journal}{Phys. Rev. B} \textbf{\bibinfo{volume}{69}},
  \bibinfo{pages}{113404} (\bibinfo{year}{2004}).

\bibitem[{\citenamefont{Csonka and Perdew}(2005)}]{Perdew-1}
\bibinfo{author}{\bibfnamefont{A.}~\bibnamefont{Csonka},
  \bibfnamefont{G.~I.~Ruzsinszky}} \bibnamefont{and}
  \bibinfo{author}{\bibfnamefont{J.~P.} \bibnamefont{Perdew}},
  \bibinfo{journal}{J. Phys. Chem. B} \textbf{\bibinfo{volume}{109}},
  \bibinfo{pages}{21471} (\bibinfo{year}{2005}).

\bibitem[{\citenamefont{Staroverov et~al.}(2003)\citenamefont{Staroverov,
  Scuseria, Tao, and Perdew}}]{Perdew-2}
\bibinfo{author}{\bibfnamefont{V.~N.} \bibnamefont{Staroverov}},
  \bibinfo{author}{\bibfnamefont{G.~E.} \bibnamefont{Scuseria}},
  \bibinfo{author}{\bibfnamefont{J.}~\bibnamefont{Tao}}, \bibnamefont{and}
  \bibinfo{author}{\bibfnamefont{J.~P.} \bibnamefont{Perdew}},
  \bibinfo{journal}{J. Chem. Phys.} \textbf{\bibinfo{volume}{119}},
  \bibinfo{pages}{12129} (\bibinfo{year}{2003}).

\bibitem[{\citenamefont{Perdew}(1986)}]{Perdew86}
\bibinfo{author}{\bibfnamefont{J.~P.} \bibnamefont{Perdew}},
  \bibinfo{journal}{Phys. Rev. B} \textbf{\bibinfo{volume}{33}},
  \bibinfo{pages}{8822} (\bibinfo{year}{1986}).

\bibitem[{\citenamefont{Becke}(1995)}]{B95}
\bibinfo{author}{\bibfnamefont{A.~D.} \bibnamefont{Becke}},
  \bibinfo{journal}{J. Chem. Phys.} \textbf{\bibinfo{volume}{104}},
  \bibinfo{pages}{1040} (\bibinfo{year}{1995}).

\bibitem[{\citenamefont{Becke}(1993{\natexlab{b}})}]{BH_and_HLYP}
\bibinfo{author}{\bibfnamefont{A.~D.} \bibnamefont{Becke}},
  \bibinfo{journal}{J. Chem. Phys.} \textbf{\bibinfo{volume}{98}},
  \bibinfo{pages}{1372} (\bibinfo{year}{1993}{\natexlab{b}}).

\bibitem[{not({\natexlab{c}})}]{note_on_cusp}
\bibinfo{note}{In DFT (and HF) the motion of a given electron is unaffected by
  the instantaneous position of the other electrons, whereas in the
  wavefunction-based approaches such as MP2 this is not the case and the short
  range electronic interactions which inevitably occur give rise to cusp
  conditions, notably the (electronic) Coulomb cusp. Such cusp conditions,
  which DFT is free of, yield wavefunctions that are exceedingly difficult to
  describe exactly with finite basis sets. See Helgaker \emph{et al.} for more
  details \cite{Helgaker_book}. Since the importance of using large basis sets
  for MP2 is clear from Fig. 2, we caution that even the largest Pople-style
  basis set, 6-311++G(3df,3pd), yields an MP2 binding energy for the water
  dimer of 230 meV/H bond, which at $\sim$15 meV from the MP2/CBS number, is
  not necessarily of sufficient accuracy to serve as a reliable benchmark.}

\bibitem[{\citenamefont{Bingel}(1967)}]{cusp_1_bingel}
\bibinfo{author}{\bibfnamefont{W.~A.} \bibnamefont{Bingel}},
  \bibinfo{journal}{Theoret. Chim. Acta (Berl.)} \textbf{\bibinfo{volume}{8}},
  \bibinfo{pages}{54} (\bibinfo{year}{1967}).

\bibitem[{\citenamefont{Kato}(1957)}]{cusp_2_kato}
\bibinfo{author}{\bibfnamefont{T.}~\bibnamefont{Kato}},
  \bibinfo{journal}{Commun. Pure Appl. Math.} \textbf{\bibinfo{volume}{10}},
  \bibinfo{pages}{151} (\bibinfo{year}{1957}).

\bibitem[{\citenamefont{Helgaker et~al.}(2004)\citenamefont{Helgaker,
  J$\o$rgensen, and Olsen}}]{Helgaker_book}
\bibinfo{author}{\bibfnamefont{T.}~\bibnamefont{Helgaker}},
  \bibinfo{author}{\bibfnamefont{P.}~\bibnamefont{J$\o$rgensen}},
  \bibnamefont{and} \bibinfo{author}{\bibfnamefont{J.}~\bibnamefont{Olsen}},
  \emph{\bibinfo{title}{Molecular Electronic-Structure Theory}}
  (\bibinfo{publisher}{John Wiley \& Sons, Inc.},
  \bibinfo{address}{Chichester}, \bibinfo{year}{2004}).

\bibitem[{\citenamefont{Xantheas}(2000)}]{Xantheas-4}
\bibinfo{author}{\bibfnamefont{S.~S.} \bibnamefont{Xantheas}},
  \bibinfo{journal}{Chem. Phys.} \textbf{\bibinfo{volume}{258}},
  \bibinfo{pages}{225} (\bibinfo{year}{2000}).

\bibitem[{\citenamefont{Ireta et~al.}(2003)\citenamefont{Ireta, Neugebauer,
  Scheffler, Rojo, and Galv\'{a}n}}]{Joel-2}
\bibinfo{author}{\bibfnamefont{J.}~\bibnamefont{Ireta}},
  \bibinfo{author}{\bibfnamefont{J.}~\bibnamefont{Neugebauer}},
  \bibinfo{author}{\bibfnamefont{M.}~\bibnamefont{Scheffler}},
  \bibinfo{author}{\bibfnamefont{A.}~\bibnamefont{Rojo}}, \bibnamefont{and}
  \bibinfo{author}{\bibfnamefont{M.}~\bibnamefont{Galv\'{a}n}},
  \bibinfo{journal}{J. Phys. Chem. B} \textbf{\bibinfo{volume}{107}},
  \bibinfo{pages}{1432} (\bibinfo{year}{2003}).

\bibitem[{\citenamefont{Gregory and Clary}(1996)}]{gregory_clary_review}
\bibinfo{author}{\bibfnamefont{J.~K.} \bibnamefont{Gregory}} \bibnamefont{and}
  \bibinfo{author}{\bibfnamefont{D.~C.} \bibnamefont{Clary}},
  \bibinfo{journal}{J. Phys. Chem.} \textbf{\bibinfo{volume}{100}},
  \bibinfo{pages}{18014} (\bibinfo{year}{1996}).

\bibitem[{\citenamefont{Mattsson et~al.}(2006)\citenamefont{Mattsson, Armiento,
  Schultz, and Mattsson}}]{Mattsson}
\bibinfo{author}{\bibfnamefont{A.~E.} \bibnamefont{Mattsson}},
  \bibinfo{author}{\bibfnamefont{R.}~\bibnamefont{Armiento}},
  \bibinfo{author}{\bibfnamefont{P.~A.} \bibnamefont{Schultz}},
  \bibnamefont{and} \bibinfo{author}{\bibfnamefont{T.~R.}
  \bibnamefont{Mattsson}}, \bibinfo{journal}{Phys. Rev. B}
  \textbf{\bibinfo{volume}{73}}, \bibinfo{pages}{195123}
  (\bibinfo{year}{2006}).

\end{thebibliography}

\end{document}